\documentclass[aip,jap,preprint,longbibliography]{revtex4-1}

\usepackage{graphicx}
\usepackage{dcolumn}
\usepackage{bm}
\usepackage{xcolor}
\usepackage{blindtext}
\usepackage{graphicx}
\usepackage{amssymb}
\usepackage{amsfonts}
\usepackage{amsmath}
\usepackage{textcomp}
\usepackage{xcolor}
\usepackage{ulem}
\usepackage{bm}
\usepackage{parskip,float,graphicx}
\usepackage[caption = false]{subfig}
\usepackage[colorlinks=true,allcolors=blue]{hyperref}
\usepackage{physics}

\begin{document}

\title{Dissipation properties of anomalous Hall effect: intrinsic vs. extrinsic magnetic materials}

\author{V. Desbuis}
\affiliation{Laboratoire des Solides Irradi\'es, CEA/DRF/IRAMIS, UMR CNRS 7642, Institut Polytechnique de Paris, Ecole Polytechnique, F-91120 Palaiseau, France}
\author{D. Lacour}
\affiliation{Université de Lorraine, CNRS, IJL, F-54000 Nancy, France}
\author{ M. Hehn}
\affiliation{Université de Lorraine, CNRS, IJL, F-54000 Nancy, France}
\author{ S. Geiskopf} 
\affiliation{Université de Lorraine, CNRS, IJL, F-54000 Nancy, France}
\author{L. Michez}
\affiliation{Aix-Marseille University, CNRS, CINaM, Marseille France}
\author{J. Rial}
\affiliation{Univ. Grenoble Alpes, CNRS, CEA, Grenoble INP, IRIG-SPINTEC, Grenoble, France}
\author{V. Baltz}
\affiliation{Université Grenoble Alpes, CNRS, CEA, Grenoble INP, IRIG-SPINTEC, Grenoble, France}
\author{J.-E. Wegrowe} \email{jean-eric.wegrowe@polytechnique.edu}
\affiliation{Laboratoire des Solides Irradi\'es, CEA/DRF/IRAMIS, UMR CNRS 7642, Institut Polytechnique de Paris, Ecole Polytechnique, F-91120 Palaiseau, France}

\date{\today}

\begin{abstract}
A comparative study of anomalous-Hall current injection and anisotropic current injection (through planar Hall effect) are studied in Hall devices contacted to a lateral load circuit. Hall currents are injected into the load circuit from three different kinds of magnetic Hall bars: $\mathrm{Mn_{5}Si_{3}}$ altermagnet, $\mathrm{Co_{75}Gd_{25}}$ ferrimagnet, and $\mathrm{Ni_{80}Fe_{20}}$ ferromagnet. The current, the voltage and the power are measured as a function of the load resistance and the Hall angle. It is observed that the power dissipated for the three kinds of materials fellow the same law as a function of load resistance and Hall angle, at the leading order in the Hall angle. Since the anomalous Hall effect in the altermagnetic Hall-bar is due to the intrinsic topological structure (i.e. due to the presence of a Berry phase in the reciprocal space), these observations suggest that the dissipative properties of anomalous Hall effect are dominated by the injection of electric charges accumulated at the edges (including electric screening), instead of the very mechanism responsible for it.
 \end{abstract}
\pacs{72.25.Mk, 85.75.-d \hfill}

\maketitle 

\section{Introduction}
The Hall current is often assumed dissipationless because it is normal to the driving field, or in other terms because it is produced by a curl-force \cite{Berry3}  (in relation to gauge invariance and geometrical phase), or again in other words, because it is due to time-reversal symmetry breaking at the microscopic scales in relation with antireciprocal Onsager relations \cite{Onsager}. In the case of electric transport in magnetic systems, both the anomalous Hall effect (AHE) \cite{AHE,Xiao}, and the planar Hall effect\cite{Goldberg,McGuire,Ritzinger,Baltz1} (PHE) (or transverse magnetoresistance) can be measured.  The former effect is described by an antisymmetric conductivity matrix, while the latter is described by a symmetric conductivity matrix. Both AHE and PHE can then be considered as archetype of Onsager reciprocity relations, because the dissipative properties of reciprocal vs. non-reciprocal relations can be studied on equal footing\cite{PRB_2024}. 

Since seventy years\cite{Luttinger}, the study of AHE has been the object of deep and fundamental theoretical developments \cite{Smit,Kondo,Berger,Nozieres}, especially since the description of the effective magnetic field generated in the reciprocal space by the Berry curvature \cite{Jungwirth,Sonin,Bruno}, following the works of Aharonov and Stern \cite{Aharonov}, or Haldane \cite{Haldane}. In the last years, the studies about AHE focused on the different possible origine of the effect, pointing-out the qualitative difference between {\it extrinsic} and {\it intrinsic} mechanisms \cite{X_Jin}. The {\it extrinsic} mechanisms for AHE being related to spin-dependent scattering (typically skew scattering and side-jump scattering), while the {\it intrinsic} mechanisms are related to non-dissipative topological phase (i.e. Berry curvature, although we note that spin-orbit coupling is always needed to generate AHE). The last property is today intensively studied in new topological materials like Heusler alloys\cite{Heusler}, topological semimetals\cite{Semimetal}, or altermagnetic materials\cite{Smerjkal,Altermag,Altermag_Bis}. In this context, the experimental characterizations are based on the comparison between the longitudinal conductivity and the Hall conductivity, measured as a function of the temperature, of the crystal structure and composition, or of a controlled density of impurities. 

\begin{figure} 
   \begin{center}
   \includegraphics[width=6.5cm]{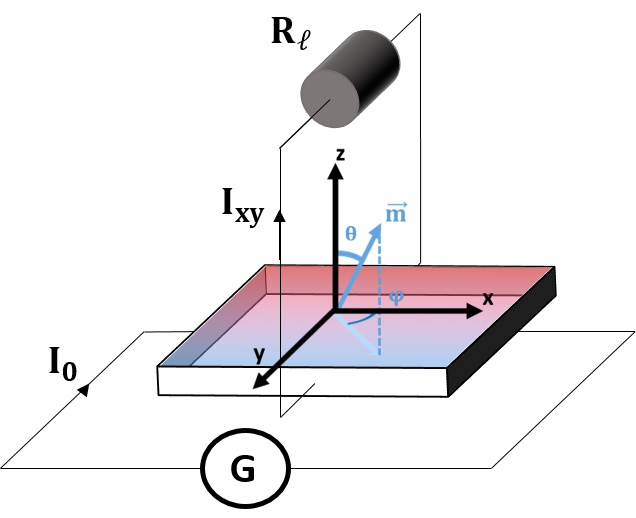} \, \, \, \, \, \, \qquad
     \includegraphics[width=5cm]{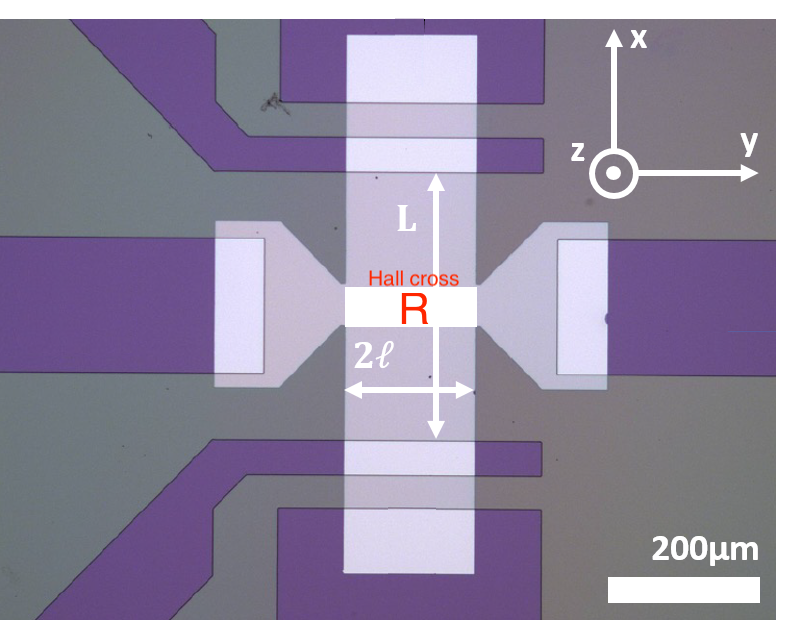}
   \end{center}
   \caption[Schema2]
{ \label{fig1}:(a) Diagram of the Hall bar contacted to a load resistance. An electric generator imposes a constant current through the longitudinal bar ($x$ direction). A transverse anomalous-Hall current is generated inside the materials (Mn$_{5}$Si$_{3}$, CoGd, and NiFe), and injected through the lateral electrodes ($y$ direction). Angles $\theta$ and $\varphi$ give the direction for the magnetization $\vec m$ (or for the magnetic order parameter for the altermagnet). (b) Optical picture of a Hall cross with the definition of width 2${\ell}$ and length L.}
\end{figure}

Yet, as shown in a recent experimental work\cite{Dan}, the electric power carried by the anomalous Hall-current can be measured directly from the Joule power dissipated in a load circuit, which is connected between the two lateral edges of the Hall bar: see Fig.\ref{fig1}.  This protocol allows an independent characterization of the AHE to be performed (the protocol has also be applied to Anomalous Nernst effect \cite{ANE}). The power dissipated on the load resistance has been shown to be approximately proportional to the square of the Hall angle, and depends on the load resistance in agreement with theoretical predictions based on non-equilibrium thermodynamics \cite{JAP3,PRB_2024}. However, the observations about Co$_{75}$Gd$_{25}$ ferrimagnetic Hall-device reported in reference\cite{Dan} were non-conclusive from the point of view of the role played by the Berry phase, because extrinsic scattering mechanisms could dominate in Co$_{75}$Gd$_{25}$, even at low temperature. 

The aim of the present report is to extend these observations to various magnetic materials, including the altermagnet Mn$_{5}$Si$_{3}$ (thin crystalline layer) that generates {\it intrinsic} AHE (i.e. related to its Berry curvature) \cite{Baltz1,Baltz2}. More precisely, the measurements performed on the altermagnetic material are compared to that performed on Co$_{75}$Gd$_{25}$ ferrimagnetic and Ni$_{80}$Fe$_{20}$ ferromagnetic Hall bars of identical geometry and measured under the same experimental protocol. The Ni$_{80}$Fe$_{20}$ (thin polycrystalline layer) plays here the role of {\it yardstick} for the conventional dissipation, as it has a strong anisotropic magnetoresistance - hence a strong PHE- together with negligible AHE. The extension of the comparative study to PHE is also highly instructive in altermagnets (as shown in the case of MnTe \cite{MnTe,Ritzinger}). 

The observations show that the profiles of the power dissipation as a function of both the load resistance $R_l$ and the Hall angle $\Theta$ are approximately superimposed (at the leading order in the Hall angle $\Theta$) for the three materials. The maximum power dissipated is approximately proportional to $\Theta^2$, and obeys the maximum power transfer theorem (maximum for the resistance matching $R=R_{\ell}$). These observations are in agreement with the phenomenological theory based on dissipative transport developed in reference \cite{PRB_2024}, that describes the lateral current injection controlled by the {\it  electric screening} present at the edges of the Hall bar \cite{JAP3,PRB_2024,JAP1,JAP2,SHE}.

\section{Phenomenological model}
The systems studied are Hall bars composed of a thin-film (defining the plane $\{x,y\}$) that are contacted to an electric generator along the longitudinal direction $x$. The usual Hall measurements consists in measuring the voltage on a voltmeter placed transversely between the two edges of the Hall bar, along the direction $y$ (open circuit configuration), while applying an external magnetic field. In magnetic materials, the magnetic field is replaced by the magnetization, which direction is defined by the radial angle $\theta$ and the azimuthal angle $\varphi$ (see Fig.\ref{fig1}). In antiferromagnets or altermagnets, the magnetization is replaced by the N\'eel vector. 

The protocol used here consists in closing the circuit by placing a load resistance $R_{\ell}$ connecting the two edges of the Hall bar, and measuring the current and the Joule power dissipated in this resistance: see Fig.\ref{fig1}. This configuration allows the dissipative properties of the Hall-current (generated by both the AHE and the PHE) to be measured directly. The corresponding theoretical analysis has been performed in recent works \cite{PRB_2024,JAP3}, and the phenomenological description is summarized below.

The local electric transport properties of a magnetic material is described by the transport equation \cite{Goldberg,McGuire}:
\begin{equation}
\vec{\mathcal E} = \rho \vec J + \Delta \rho  \left ( \vec m \cdot \vec J \right ) \vec m + \rho_{xy}\, \vec{m} \times \vec J
\label{Ohm}
\end{equation}
 where $\vec {\mathcal E}$ is the electric field, $\vec J$ the electric current density, and $\vec m$ is the unit vector the gives the direction of the magnetization (assumed uniform), $\rho$ is the longitudinal resistivity, and $\rho_{xy}$ is the anomalous Hall coefficient, that couples the two directions of the space in the plane of the Hall bar. If the magnetization is not totally saturated, an analysis of the magnetization states is necessary \cite{Dan}. 
Note that the energy of the carriers also contains the contribution due to the electric charge accumulation and screening, so that the effective electric field $\vec{\mathcal E}$ should be generalized with the gradient of the chemical potential \cite{Rq_mu}.
 
In a first step we assume homogeneous and isotropic materials.  In addition, we assume that the electric current flowing in the $xy$-plane only. The Ohm's law Eq.(\ref{Ohm}) can then be written $\vec {\mathcal E}= \hat \rho \vec J$. The resistivity matrix $\hat \rho$ can be reduced to the $2\times 2$ matrix in an arbitrary orthonormal $2D$ frame:
\begin{align}
  \begin{aligned}[c]
    \hat{\rho}= \rho \left(\begin{array}{cc}
      1 + \frac{\Delta \rho}{\rho} m_x^2 \, & \, \, \Theta_{pl} + \Theta_{an}\\
    \Theta_{pl} - \Theta_{an}  \, & \,  \, 1 +  \frac{\Delta \rho}{\rho} m_y^2 \\
    \end{array}\right)
      \label{Matrix2}
  \end{aligned}
\end{align}
where the diagonal term $\frac{\Delta \rho}{\rho} m_x^2 \equiv \frac{\Delta \rho}{\rho} sin^2(\varphi) sin^2(\theta)$ is the anisotropic magnetoresistance (AMR) that is measured longitudinally, along the $x$ axis. Note that the AMR is small with respect to unity ($ \frac{\Delta \rho}{\rho} \ll 1$), and we will not consider the AMR contribution in the following.
More interesting for the present study are the two {\it Hall angles}\cite{Rqe}, defined as (see details in reference \cite{PRB_2024}):
\begin{align}
  \begin{aligned}[c]
    \Theta_{pl} =   \frac{ \Delta \rho}{2\rho} \, \sin^2(\theta) \, \sin(2 \varphi)
  \end{aligned}
  \qquad \text{ and } \qquad
  \begin{aligned}[c]
    \Theta_{an} = - \frac{\rho_{xy}}{\rho} \, \cos(\theta)
  \end{aligned}
  \label{Hall_Angle}
\end{align} 
where the index $pl$ stands for PHE and the index $an$ stands for AHE. Note that the material is characterized by the maximum values $\Theta_{max}^{pl} =  \frac{ \Delta \rho}{2\rho}$ reached for $\theta= \pi/2 $, $\varphi = \pi/4$, and $\Theta_{max}^{an} =  \frac{\rho_{xy}}{\rho}$ reached for $\theta = 0$.

 In the case of an {\it open circuit} ($R_{load} \rightarrow \infty$) the transverse current density $J_y$ vanishes\cite{JAP1} and the Hall voltage measured between the two lateral edges is, after integration of $\mathcal E_y$ over the length of the Hall bar:
\begin{equation}
V^{0}_{xy}(\theta,\varphi) = 2 \ell \, \rho I_0 \left (    \Theta_{pl}(\theta,\varphi) +  \Theta_{an}(\theta) \right ) 
\label{Hall_Voltage_Open}
\end{equation}

where $V^{0}_{xy} = V^0(y=-\ell) - V^0(y=\ell)$ and the superscript $V^0$ stands for the open-circuit condition. From the experimental point-of-view,  the relations Eqs.(\ref{Hall_Angle}) allows the two effects, PHE and AHE, to be characterized  in a univocal way while measuring both the voltage and the current in the load circuit. 
This expression of the Hall voltage justify the use of the term PHE, i.e. the reference to {\it Hall} despite the fact that the effect is just the transverse anisotropic magnetoresistance. 
We are aware that the name {\it planar Hall angle} could be problematic, and it was criticized since the first publication \cite{Goldberg}. The criticism is justified: the PHE is fundamentally different from Hall effects because it conserves {\it the time reversal symmetry} at the microscopic scale, while AHE breaks this symmetry. This striking microscopic property is precisely reflected at the macroscopic scale by the Onsager reciprocity relations of the first\cite{Onsager1} kind (PHE), and of the second\cite{Onsager} kind (AHE) through the dichotomy between the symmetrical part (PHE) and the antisymmetrical part (AHE) of the conductivity matrix, as shown in Eq.(\ref{Hall_Angle}). However, as will be shown in the following, this fundamental difference is not reflected in a striking way at the level of the Hall-current measured in a load circuit.

Due to the accumulation of electric charges at the edges, and the resulting non-equilibrium screening effect, the problem is not entirely defined by the transport equation Eq.(\ref{Ohm}) and the continuity equation, that are local. For this reason, a specific variational approach has been developed\cite{PRB_2024} in order to define the stationary states. The method is based on the minimization of the power dissipated in the system under the constrains imposed by the generator and the circuit's configuration. The power density takes the usual form: $p_J = \vec J. \vec{\mathcal E}$, and the total power is obtained after integrating $p_J$ over the whole system, including the load circuit. The resulting Hall-voltage at the load terminals is given by the expressions: 
\begin{align}
  \begin{aligned}[c]
V^{pl}_{xy} = 2 \ell \rho J_x^0 \, \frac{1}{1+\alpha} \,  \frac{\Theta_{pl}}{1-\Theta_{pl}^2} 
 \end{aligned}
  \qquad \text{ and } \qquad
  \begin{aligned}[c]
V^{an}_{xy} =2 \ell \rho J_x^0 \, \frac{1}{1+\alpha} \,  \frac{\Theta_{an}}{1+\Theta_{an}^2} 
  \end{aligned}
 \label{Hall_Voltage}
\end{align}
where the parameter $\alpha = R/R_{\ell}$ is the ratio of the resistance of the transverse part of the Hall bar over the load resistance $R_{\ell}$, and $J_x^0$ is the current injected by the generator at $\Theta = 0$. Beside the specific behaviour of $\Theta(\theta,\varphi)$ shown in Eq.(\ref{Hall_Angle}), the only difference between AHE and PHE is contained in the sign $\pm$ that appears in the expression $1/\left (1 \pm \Theta^2 \right )$.  \\

On the other hand, the Joule power $P=V_{xy}.I_{xy}$ dissipated at the load terminals by the anomalous or planar Hall-currents injection reads: 
\begin{align}
  \begin{aligned}[c]
P(\Theta_{pl},\alpha)=P_{0} \, \frac{\alpha}{(1+\alpha)^{2}} \, \frac{\Theta_{pl}^{2}}{1-\Theta_{pl}^{2}},
 \end{aligned}
  \qquad \text{ and } \qquad
  \begin{aligned}[c]
P(\Theta_{an},\alpha)=P_{0} \, \frac{\alpha}{(1+\alpha)^{2}} \, \frac{\Theta_{an}^{2}}{1+\Theta_{an}^{2}} ,
  \end{aligned}
 \label{Power}
\end{align} 

where P$_{0}$ is the power injected into the Hall bar for open circuit (or for $\Theta = 0$). As expected, the PHE dissipates more than the AHE. 
 
 The objective of the two next sections is to confirm experimentally the validity and the ``universality'' of the expressions Eq.(\ref{Hall_Voltage}) and Eq.(\ref{Power}) for the three specific materials. 

\section{Transverse Voltage for open circuit configuration}
The characterization of the voltage $V^{0}_{xy}(\theta,\varphi)$ for the open-circuit configuration is of fundamental importance in the present study, because we are exclusively interested in the currents generated by AHE or PHE injected into the lateral circuit. This means that two main artifacts has to be corrected that are (i) the thermoelectric contributions and (ii) the effect of the misalignement  (of some hundreds of nanometers) of the two lateral contacts of the Hall cross. The first artifact (i) is corrected by a shift in the I(V) curve (removing the non-zero voltage measured when the generator is switched off), while the second artifact (ii) is corrected by removing the offset voltage from the angular dependence of $V_{xy}(\theta,\varphi)$ (see figures below). 

The ferromagnetic material Ni$_{80}$Fe$_{20}$ (permalloy) has been chosen because of its strong PHE and negligible AHE ($\Delta\rho \gg \rho_{H}$) \cite{Madon}, while the ferrimagnetic alloy Co$_{75}$Gd$_{25}$ has been chosen because of its strong AHE and negligible PHE ($\rho_{H}\gg\Delta\rho$). Both materials are measured at room temperature. In contrast, the altermagnet Mn$_{5}$Si$_{3}$ has been chosen because the AHE is generated by an intrinsic mechanism\cite{Baltz2}, due to the presence of a Berry curvature in the reciprocal space (together with spin-orbit coupling). The experiments on Mn$_{5}$Si$_{3}$  are performed at 70K, temperature at which the material exhibits the altermagnetic phase with the presence of spontaneous AHE and vanishing net magnetization\cite{Baltz1}. The out-of-plane signal $V_{xy}(\theta)$ (for open circuit configuration) is due to AHE, and a detailed description of the sample can be found in reference\cite{Baltz2}. 

The experimental protocol for the voltage measurements $V_{xy}(\theta,\varphi)$ at the open-circuit configuration is the following. If the magnetization is saturated by the external field $\vec H$, the direction of the magnetization $\vec m$ coincides with the direction of the vector $\vec H$. In that case, the magnetization state is well-known, and the voltage is a function of the angles $\theta$ and $\varphi$. The contributions of AHE and PHE to the voltage $V_{xy}$ can then be directly deduced from Eqs.(\ref{Hall_Voltage_Open}). This is the case for the measurements of Ni$_{80}$Fe$_{20}$ with the magnetic field in the plane. 
 If the magnetization is not fully saturated, a correction should be apply, from micromagnetic considerations.
\begin{figure}[H]
\centering
\includegraphics[width=15cm]{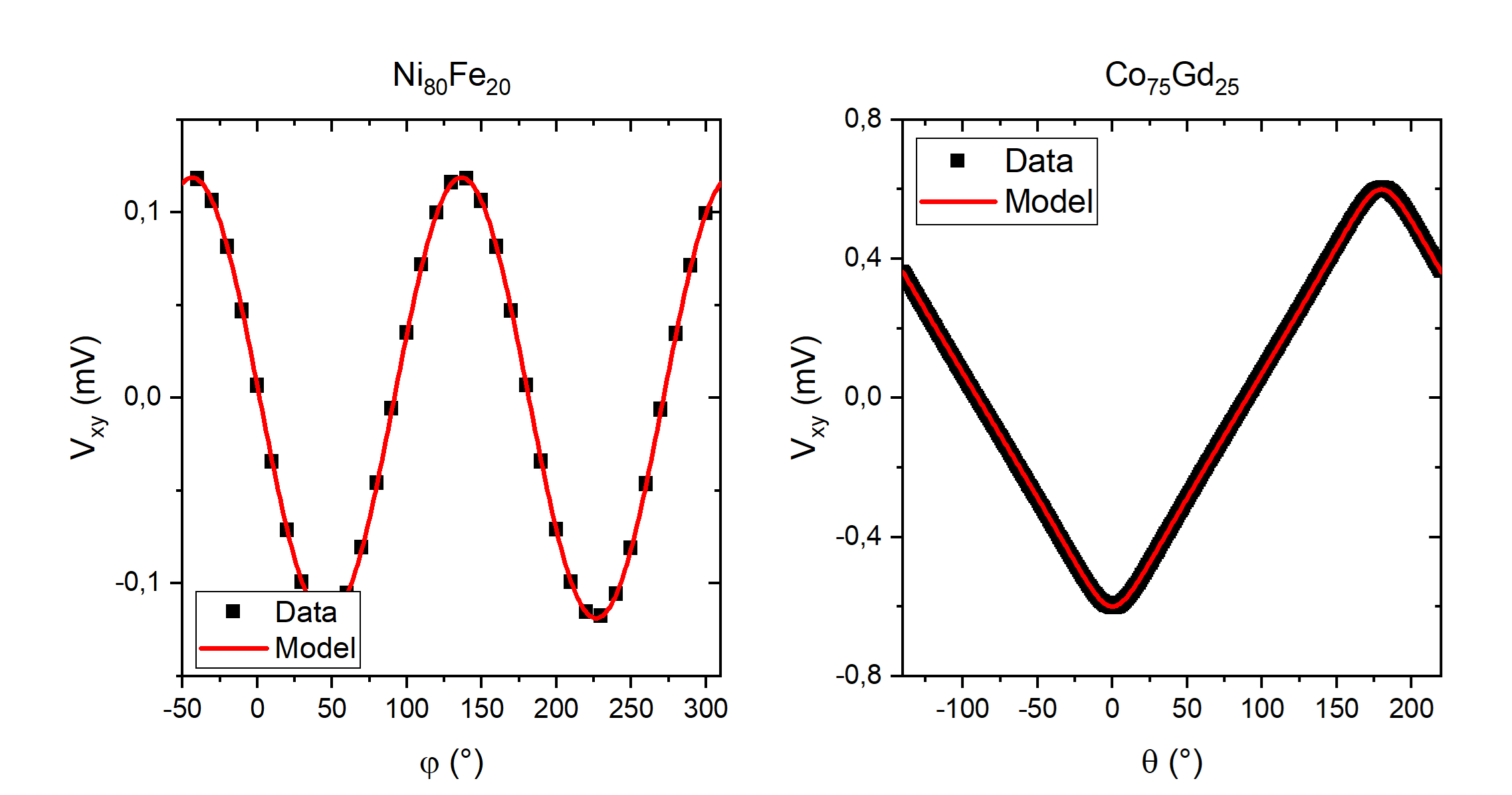}
\caption{\label{NiFe} Left: Planar Hall voltages in Ni$_{80}$Fe$_{20}$ versus in-the plane azimuthal angle $\varphi$ of the applied field $H_a$ at $H_a=1.5$ T. Right Anomalous Hall voltages in GdCo versus out-of-plane radial angle $\theta$ of the applied field $H_a$ at $H_a=1.5$ T. The magnetization of Co$_{75}$Gd$_{25}$ is not fully saturated at $1.2$ T. The model is the result of the calculation of the equilibrium states of the magnetization for each orientation $\theta$ of the applied field (see reference \cite{Dan}).}
\end{figure}
\begin{figure}[H]
\centering
\includegraphics[width=7.5cm]{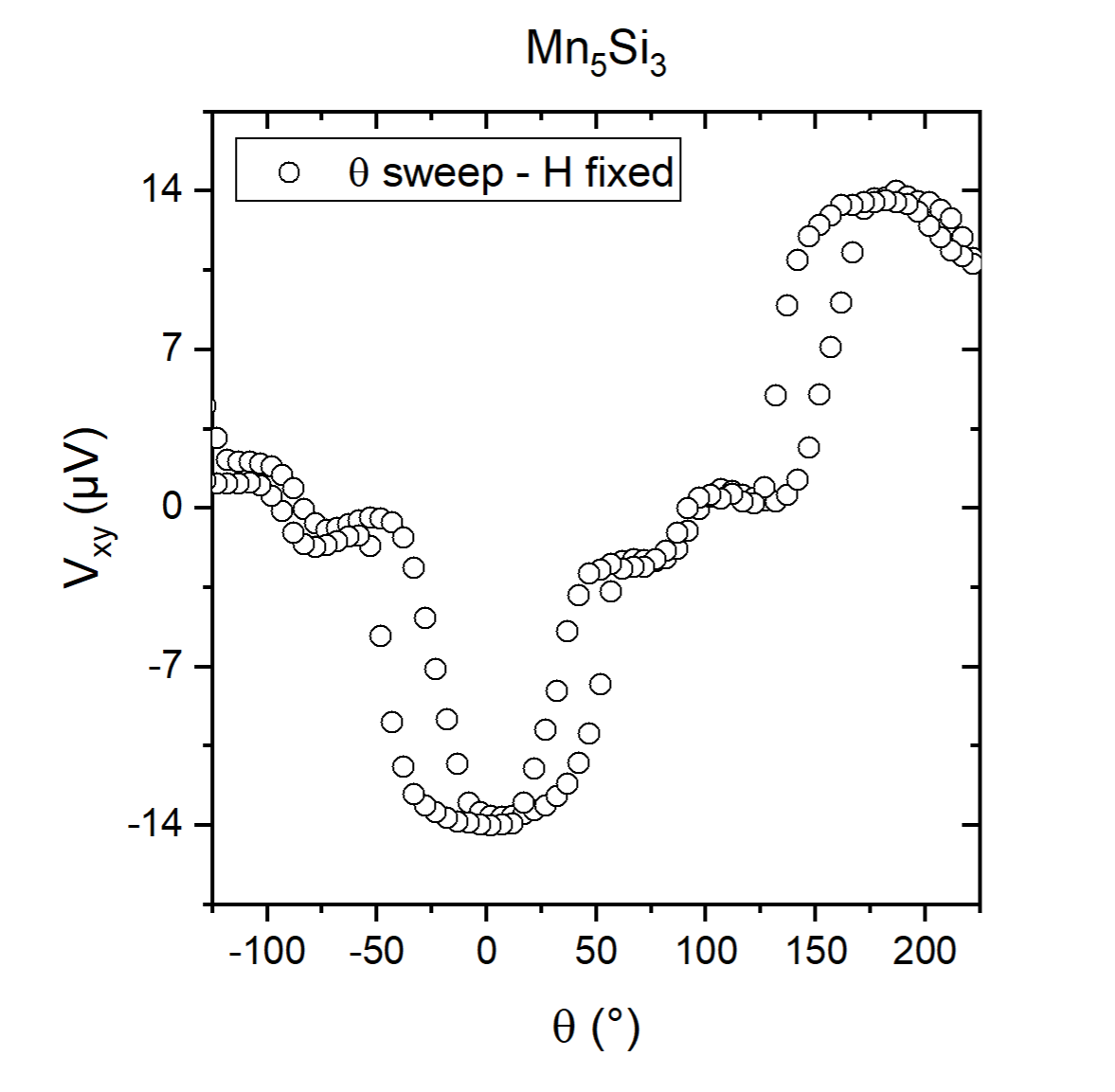}
\caption{\label{Mn5Si3} Hall voltages in Mn$_{5}$Si$_{3}$ versus out-of-plane angle $\theta$ for an applied field $H_{a} = 1.2$T (white circles), measured at T=70K. The current is injected along the $[2 \overline 1 \overline 1 0]$ direction \cite{Baltz2}.}
\end{figure}
Indeed, for Co$_{75}$Gd$_{25}$, we see that the voltage does not strictly follows the AHE profile proportional to $cos(\theta_H)$, where $\theta_H$ is the angle of the applied field. However, after having corrected the effect of non-saturation, the calculated profile (see Eq.(\ref{Hall_Angle}) and Eq.(\ref{Hall_Voltage_Open})), coincides with the measurements (as shown in Fig.\ref{NiFe}), thus indicating a $cos(\theta)$ behavior. The relation between $\theta$ and the angle of the applied field $\theta_H$ has been calculated numerically, as described in reference \cite{Dan}.
In the case of the altermagnet Mn$_{5}$Si$_{3}$, the problem is more subtil because these is no real saturation (the magnetization is nearly zero but the order parameter depends on the magnetic field), and minor hysteresis loops can be seen in Fig.\ref{Mn5Si3}. However, as for  the Co$_{75}$Gd$_{25}$ sample, the fact that the magnetization state is not well-known is not an obstacle for the application of our measurement protocol (see below).
The maximum values of the anomalous or planar Hall angles are reached for $\Theta_{max}^{CoGd} \approx 5.10^{-3}$ for Co$_{75}$Gd$_{25}$ \cite{Madon,Dan}, $\Theta_{max}^{NiFe} \approx 8 . 10^{-3}$ for Ni$_{80}$Fe$_{20}$ for the permalloy Hall bar, and $\Theta_{max}^{Mn_{5}Si_{3}} \approx  3. 10^{-5}$ for the altermagnet. These quantities will be used below for the comparative study.

Note that for all three materials, the voltage $V_{xy}(\theta, \varphi)$ plotted in Fig.\ref{NiFe} and Fig.\ref{Mn5Si3} are centered at zero (or in other terms, the mean value of the function $\Theta(\theta,\varphi)$ is zero). However, an offset (i.e. a constant shift of the function $\Theta(\theta,\varphi)$), due mainly to thermoelectric effects, has been removed). 

\section{Power efficiency}
When the lateral edges are contacted to a load resistance the circuit is closed and a Hall current I$_{xy}(\theta, \varphi)$ is flowing into the load. This current has the same profile as the voltage for the open circuit, i.e. the profiles follow the relation Eq.(\ref{Hall_Angle}) and Eq.(\ref{Hall_Voltage_Open}), as shown in a previous work \cite{Dan}. The measurements of the Hall current as a function of the angle ($ \theta$ or $\varphi$) are performed for each value of the load resistances, tuned from 1$\Omega$ up to 12k$\Omega$ in the case of Mn$_{5}$Si$_{3}$. 

\begin{figure}[H]
\centering
\includegraphics[width=8.5cm]{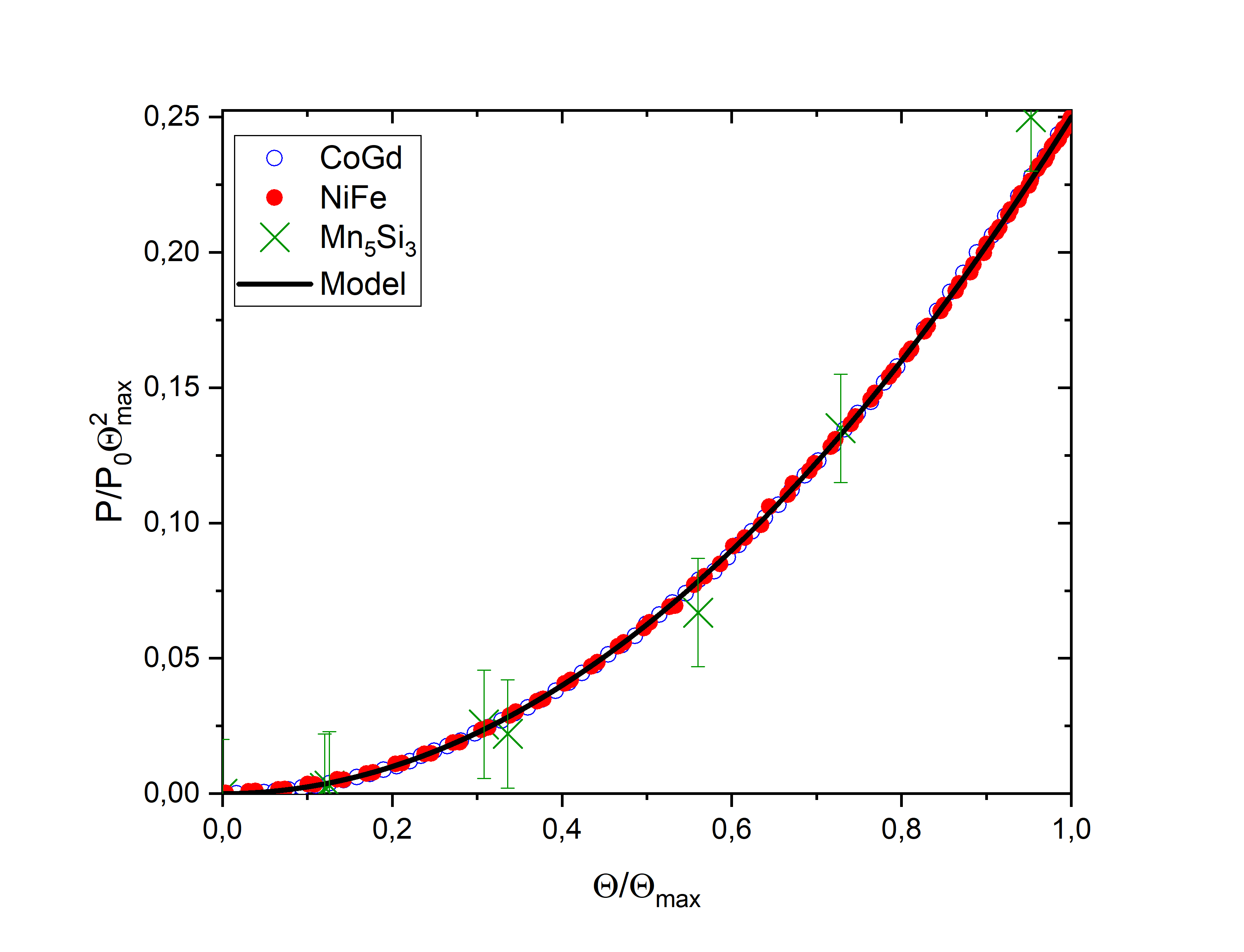}
\includegraphics[width=7cm]{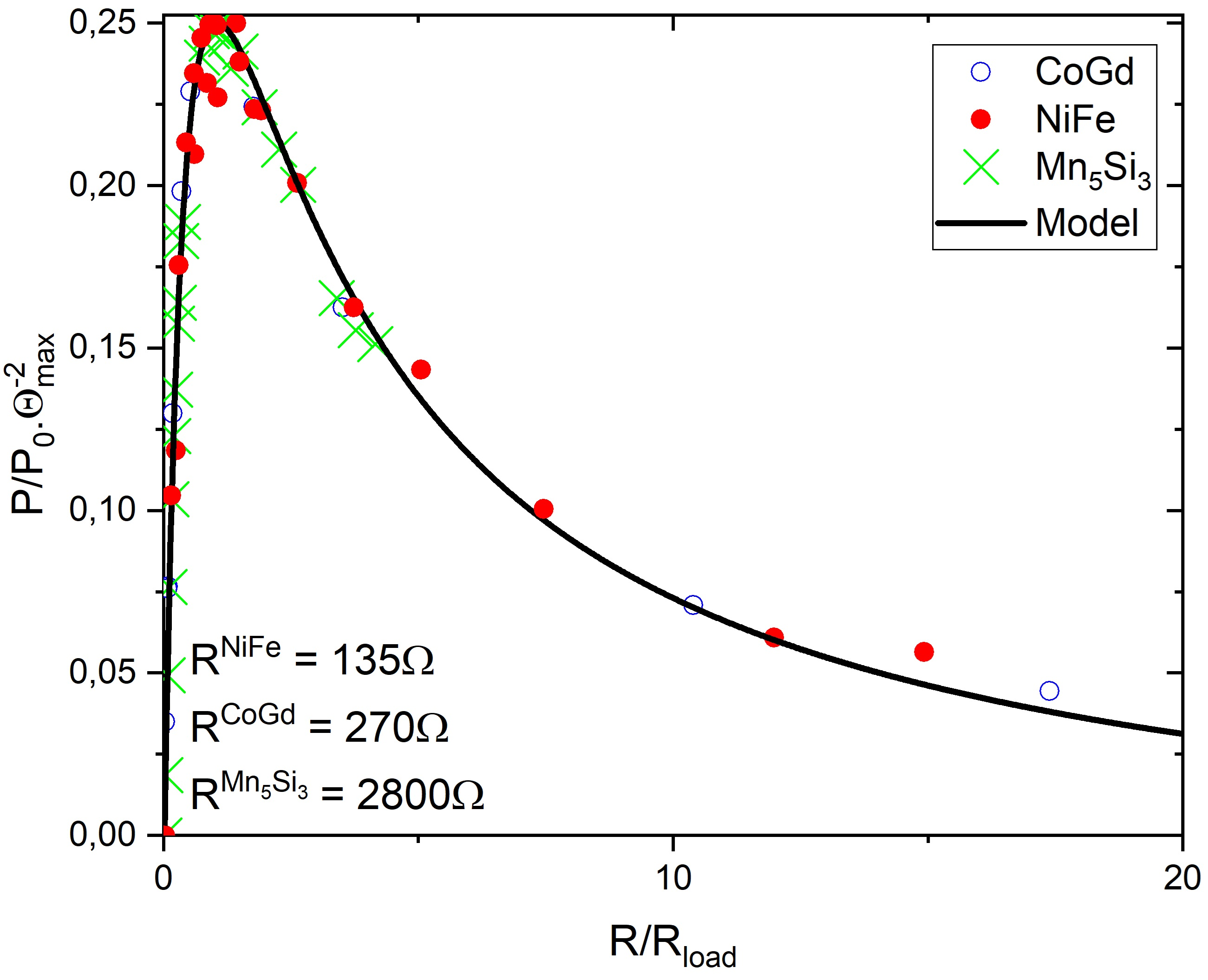}
\caption{\label{NormP} Normalized dissipated power in the load resistance for the three materials as a function of Normalized Hall angle (left) and Normalized load resistance (right). Continuous black line are deduced from the model discussed in the text. The error bars on the left indicate the uncertainty due to the hysteresis shown in Fig.3.}
\end{figure}

The Hall current flowing into the lateral circuit dissipates the power $P(\theta, \varphi)= V_{xy}.I_{xy}$. The profiles of the power extracted from Hall currents in the three materials (NiFe, CoGd and Mn$_{5}$Si$_{3}$)) are plotted, after being normalized by the power $P_0 = V_{xx}I_{xx}$ (constant for a given sample) injected by the generator in the open circuit configuration. Furthermore, the load resistance $R_\ell$ is normalized with the resistance $R$ of the material, corresponding to region defined by the cross of the Hall device (see Fig.1 right), through which the lateral Hall-current is flowing (as studied in reference \cite{Dan}). The parameter $R$ can be measured with injecting a known current along the transverse direction in the Hall cross.  The values obtained are $R^{CoGd} = 270$ $\Omega$ for Co$_{75}$Gd$_{25}$, $R^{NiFe} = 135$ $\Omega$ for permalloy, and $R^{Mn_{5}Si_{3}} = 2800$ $\Omega$ for the altermagnet. The results about normalized power dissipation in the load are shown on Fig.\ref{NormP}. 
The left plot of Fig.\ref{NormP} shows the dependence of the normalized dissipated power $P(\theta,\varphi)/\left ( P_0 \Theta_{max}^2 \right )$ as a function of $\Theta(\theta,\varphi)/\Theta_{max}$, and the right plot of Fig.\ref{NormP} shows the normalized dissipated power as a function of the ratio $R/R_{\ell}$. The first important observation is that the power is of the order of  $\Theta_{max}^2$, i.e. it is a factor $10^{-4}$ to $10^{-9}$ of the power injected from the generator. From this point of view, PHE and AHE (intrinsic and extrinsic) are indeed dissipative, but the resulting dissipation is small.

The second important observation is that the normalized power profiles can be superimposed at the leading order of $\Theta$ as seen on Fig.\ref{NormP}. The correction that allows PHE to be differentiated from AHE is contained in the factor $1/\left ( 1 \pm \Theta^{2} \right )$ in Eq.(\ref{Hall_Voltage}). Considering that Hall angles for all materials are below 10$^{-2}$, the factor between the two extreme cases AHE ($1/\left ( 1 + \Theta^{2} \right )$) and PHE ($1/\left ( 1 - \Theta^{2} \right )$) does not exceed 10$^{-4}$, compared to $1$. This correction can only be seen with a resolution of the next to leading order in $\Theta$. In the present work (which is a first comparative study based on this experimental protocol), the next to leading order resolution has not been reached, and the discussion will then focus on the similarity of the three profiles at the leading order for all three materials.

A third observation can be made on Fig.\ref{NormP}: the normalized power profiles can also be superimposed as a function of $\frac{R}{R_{\ell}}$ and all measured points follow the curve given by Eq.(\ref{Power}). The maximum of the normalized power equals 0.25 and is reached for the same value $\frac{R}{R_{\ell}}=1$ for all three materials. The results shown on Fig.\ref{NormP} demonstrates the ``universality'' of the power dissipation of Hall currents for the three specific materials studied. More precisely, the dissipated power do not depend on the mechanisms involved, nor on the materials used. These very properties of the different mechanism for AHE and anisotropic magnetoresistance are contained in the two parameters $\Theta_{max}$ and the resistivity of the material through the resistance $R$.

\section{Conclusion}
 
We have studied the effect of the transverse current injection and the corresponding dissipation into a load circuit, due to PHE and AHE in three different materials. The materials have been chosen in order to compare Hall-like currents generated by very different mechanisms. 
Indeed, the Co$_{75}$Gd$_{25}$ ferrimagnetic material shows a typical AHE due to extrinsic effect, while the altermagnet Mn$_{5}$Si$_{3}$ shows spontaneous AHE, the amplitude of which is due to intrinsic properties related to Berry curvature in the reciprocal space (when spin-orbit coupling still triggers the effect). On the other hand, the Ni$_{80}$Fe$_{20}$ material shows a typical PHE effect (with negligible AHE) due to s-d scattering.

The first observation shows that amplitude of the power is typically of the order of the square of the Hall angle $\Theta^2$ in all cases. This is a small quantity in general, varying form $10^{-4}$ to $10^{-9}$ fraction of the power $P_0$ injected from the generator in our case (some exceptions of large Hall angles have however been reportes, e.g. measured in topological semimetals\cite{Parkin}). 
The second observation-  presented in Eq.(\ref{Power}) - shows that the curves of the normalized power are superimposed for the three materials, at the leading order in $\Theta$. 

This behavior is predicted in a phenomenological model based on the Onsager reciprocity relations, that takes into account the electric screening effect occurring at the edges\cite{JAP3,PRB_2024}. This shows that the nature of the microscopic mechanisms responsible for the different kinds of Hall effects (intrinsic topological properties, extrinsic scattering, etc) do not play a crucial role for the phenomenological electric properties of the Hall devices. The main physical picture, at macroscopic scale, is the accumulation of the electric charges at the edges, that are injected into the load.

This study is being pursued in order to reach the next to leading order in the Hall angle $\Theta$. More importantly, the resistance matching property shown in the present work should find an application for optimization of spin-orbit torque (SOT) devices. Beyond, from a more fundamental viewpoint, the characteristics proper to altermagnets\cite{MnTe, Gomoney} should also be observed through the power dissipated in the load circuit at the next to leading order. An appropriate description would then be necessary, that goes beyond the basic phenomenological model proposed here.  

\section{data availability statement} 
The data that support the findings of this study are available from the corresponding author upon reasonable request.

\begin{acknowledgements}
The groups of the CINaM and SPINTEC is supported by French national research agency (ANR-22-EXSP-0007) and the Deutsche Forschungsgemeinschaft (DFG), Germany (Project HEXAS - Grant No. ANR-24-CE92-0038-02). The group of IJL is supported by the interdisciplinary project LUE MAT-PULSE, part of the French PIA project {\it Lorraine Universit\'e d'Excellence} reference ANR-15-IDEX-04-LUE and by the ANR through the France 2030 government grants PEPR SPIN SPINMAT ANR-22-EXSP-0007.
\end{acknowledgements}
\newpage
\pagebreak


\begin{thebibliography}{47}
\bibitem{Rq_mu} In other terms, the electric field is not only defined d by the electric potential, but also by the gradient of the density of electric carriers (i.e. both drift and diffusion should be  taken into account).
\bibitem{Rqe} Our definition of the parameter $\Theta$  is in fact the tangent of the usual Hall angle $\Theta_{us}$:  $tan\left (\Theta \right ) = \Theta_{us}$. The two definitions are very close, since  $\Theta$ is small ($\Theta < 0.01$ in our case).
\end{thebibliography}
\end{document}